\documentclass[aps,prl,twocolumn,amsfonts,showpacs]{revtex4-1}
\usepackage{epsfig,amsopn,amssymb,amsmath,amsfonts}
\usepackage{graphicx}
\usepackage{wasysym}
\usepackage{latexsym}
\newcommand{\gae}{\lower 2pt \hbox{$\, \buildrel {\scriptstyle >}\over {\scriptstyle
\sim}\,$}}
\newcommand{\lae}{\lower 2pt \hbox{$\, \buildrel {\scriptstyle <}\over {\scriptstyle
\sim}\,$}}

\begin{document}

\def\be{\begin{equation}}
\def\ee{\end{equation}}
\def\bea{\begin{eqnarray}}
\def\eea{\end{eqnarray}}
\def\bef{\begin{figure}}
\def\eef{\end{figure}}
\def\l{\label}
\def\fr{\frac}
\def\th{\theta}
\def\o{\omega}
\def\O{\Omega}
\def\eps{\epsilon}
\def\p{\partial}
\title{Long-range steady state density profiles induced by localized drive}
\author{Tridib Sadhu$^\dagger$, Satya N. Majumdar$^\S$, and David Mukamel$^\dagger$}
\affiliation{
{\small $^\dagger$Physics of Complex Systems, Weizmann Institute of Science, Rehovot 76100, Israel.}\\
\small $^\S$Univ. Paris-Sud, CNRS, LPTMS, UMR 8626, Orsay F-01405,
France.}
\date{\today}
\begin{abstract}
We show that the presence of a localized drive in an otherwise diffusive system
results in steady-state density and current profiles that decay
algebraically to their global average value, away from the drive in two or higher dimensions. An analogy to
an electrostatic
problem is established, whereby the density profile induced by a
driving bond maps onto the electrostatic potential due to an electric
dipole located along the bond. The dipole strength is proportional
to the drive, and is determined self-consistently by solving the
electrostatic problem. The profile resulting from a localized
configuration of more than one driving bond can be straightforwardly
determined by the superposition principle of electrostatics. This
picture is shown to hold even in the presence of exclusion
interaction between particles.
\end{abstract}
\maketitle

The effect of a local perturbation on the steady state density
profile of systems of interacting particles has been studied in a
wide variety of contexts. When a system under thermal equilibrium
conditions is perturbed by a localized external potential, the
equilibrium density is changed only {\it locally} under generic
conditions. This is a result of the fact that as long as the system
is not at a critical point, it is characterized by a finite
correlation length. This is not necessarily the case in driven
systems where detailed balance is not satisfied \cite{ZIA}.
Algebraically decaying correlations in the steady state profiles have
been found in a number of {\em boundary driven} models such as the
Asymmetric Simple Exclusion Process (ASEP) \cite{BLYTH}, interface
growth models \cite{KM} and transport models in one and higher
dimensions \cite{LIVI,DHAR,SPOHN}.

A natural question is what happens when the drive is localized not on the
boundaries but in the {\em bulk}.
In a recently studied model of a
one-dimensional Symmetric Simple Exclusion Process (SSEP) it was
shown that the presence of a single driving bond (battery) in the
bulk {\it does not generate} any algebraic density profile in  the steady
state \cite{BDL}. The density profile was found to be flat away
from the battery, albeit with a discontinuity at the location
of the battery.

In the present Letter we consider the effect of driving bonds localized in a finite region in an infinitely large many body system.
We demonstrate
that under rather generic conditions a localized drive in an
otherwise equilibrium system in dimensions higher than one, results
in a steady state density profile with an algebraically decaying tail. This is
done by first studying the case of non-interacting particles
diffusing on a d-dimensional lattice with a directional drive along
a single bond (battery). We then generalize the results to arbitrary
localized configurations of driving bonds, and to the case of
particles with exclusion interaction.

In the case of non-interacting particles with a single driving bond,
we show that the density profile can be mapped onto the
electrostatic potential generated by an electric dipole located at
the driving bond, whose strength can be calculated
self-consistently. Thus, for example, in $d=2$ dimensions, the
density profile decays as $1/r$ at distance $r$ away from the
driving bond, in all directions except the one perpendicular to the
drive, where it decays as $1/r^{2}$. More interestingly, other
localized configurations of driving bonds result in different
power-law profiles. In this case the density profile can be
determined by a linear superposition of the profiles generated by
each driving bond. For example, when the electric dipoles
corresponding to two driving bonds form a quadrupole, the density
profile generically decays as $1/r^2$ while in some specific
directions it decays as $1/r^4$, at large distances. The correspondence to
the electrostatic problem still holds when {\it local} exclusion is switched on. The only
difference is in the dipole strength which, unlike the
noninteracting case, can not be determined self-consistently. In
the interacting case, our results thus generalize the one
dimensional situation studied in \cite{BDL} and show that for $d\ge
2$, the density profile decays algebraically away from the battery.

\begin{figure}
\includegraphics[width=0.9\hsize]{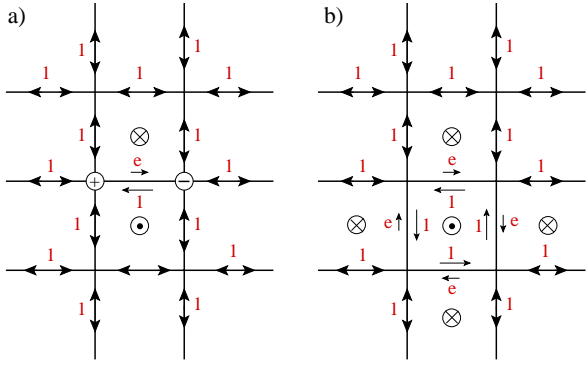}
\caption{(a) Bond-configuration with a single driving bond $(0,0)\rightarrow
(1,0)$ and hopping rate $e=(1-\epsilon)$. Everywhere else the rate
is $1$. The $\oplus$ and $\ominus$ symbols denote the sign of the
electric charge. The symbol $\odot$ represents the magnetic flux coming
out of the plane, and $\otimes$ represents the flux going in.
(b) A configuration of closed loop of driving bonds generating no
electric charges. The magnetic flux through the central plaquette is
four times larger than that of the other four neighboring
plaquettes.}
\label{fig1}
\end{figure}

We start with the simple case of non-interacting particles diffusing
in a medium with a single driving bond. As an illustration we
consider explicitly the $d=2$ case. Generalization to arbitrary
dimensions is straightforward. We consider a two-dimensional square
lattice of $L\times L$ sites with periodic boundary
conditions. There are $N= \rho V$ noninteracting particles where $V=L^2$
is the number of sites and $\rho$ is the global average density of
particles. Each particle performs an independent random walk in
continuous time. A particle at site $\vec r\equiv(m,n)\ne (0,0)$ can
hop to any of the neighboring sites with rate $1$. We introduce a
localized drive by setting the hopping rate across the bond
$(0,0)\to (1,0)$ to be $(1-\epsilon)$ with $\epsilon \le 1$ (see
Fig.\ref{fig1}a). In the absence of the localized drive
($\epsilon=0$), detailed balance holds and the system reaches a
steady state with a flat density profile with density $\rho$ at each
site. When the localized drive $\epsilon$ is switched on, it
manifestly violates the detailed balance and induces a global
modification of the steady state density profile.

Let $\phi(m,n,t)$ denote the average density of particles at site
$(m,n)$ at time $t$. Its time evolution can be easily written down
by counting the incoming and outgoing moves from each site. For sites
$(m,n)\ne \{(0,0), (1,0)\}$, it is easy to see $\phi(m,n,t)$
satisfies the standard diffusion equation, $\partial_t \phi(m,n,t)=
\nabla^2 \phi(m,n,t)$, where $\nabla^2$ is the discrete Laplacian:
$\nabla^2 \phi(m,n)=
\phi(m+1,n)+\phi(m-1,n)+\phi(m,n+1)+\phi(m,n-1)-4\phi(m,n)$. The
evolution equations are slightly different for the two special sites
$(0,0)$ and $(1,0)$ connecting the driving bond: $\partial_t
\phi(0,0,t)= \nabla^2 \phi(0,0,t) +\epsilon \phi(0,0,t)$ and
$\partial_t \phi(1,0,t)=  \nabla^2 \phi(1,0,t) -\epsilon
\phi(0,0,t)$. It is useful to write these evolution equations in a
combined form by introducing Kronecker delta symbols. For brevity, let
us also denote ${\vec r}=(m,n)$, ${\vec 0}=(0,0)$ and ${\vec
e_1}=(1,0)$. In the long time limit the system approaches a time
independent steady state $\phi(\vec r)$ satisfying
\begin{equation}
\nabla^2 \phi(\vec r) =-\epsilon \phi(\vec 0)\left[\delta_{\vec r,\vec 0}-\delta_{\vec r,\vec e_1}\right]
\label{diff1}
\end{equation}
for all $\vec r$. Here $\delta_{\vec r,\vec r'}$ is the Kronecker
delta function. It is instructive to first note the formal
resemblance of \eqref{diff1} with the Poisson equation (lattice
version) of the $2$-d electrostatic problem. One identifies
$\phi(\vec r)$ as the electrostatic potential and the right hand
side (rhs) of \eqref{diff1} is identified with two point charges of
equal strength but of opposite signs sitting respectively at the two
ends of the driving bond. Thus we have effectively a dipole sitting
on the weak bond. However, unlike in standard electrostatics the
charge strength $\epsilon \phi(\vec 0)$ has to be determined
self-consistently.

We consider the thermodynamic limit $L\to \infty$, $N\to \infty$
with density per site $\rho=N/V$ fixed. The exact solution of
\eqref{diff1} at any $\vec r$ can be expressed in terms of the
lattice Green's function $G(\vec r, {\vec r_0})$ which is the Coulomb
potential
due to a single point charge of unit strength at $\vec r_0$, that
satisfies $\nabla^2 G= - \delta_{\vec r, {\vec r_0}}$. Using the
superposition principle one obtains the solution
\begin{equation}
\phi(\vec r)= \rho + \epsilon \phi(\vec 0)\left[G(\vec r, \vec
0)-G(\vec r, \vec e_1)\right].
\label{dens1}
\end{equation}
The constant $\phi(\vec 0)$ can be determined self-consistently by
substituting $\vec r=\vec 0$ in \eqref{dens1}. This gives,
$\phi(\vec 0)= \rho/[1-\epsilon (G(\vec 0, \vec 0)-G(\vec 0, \vec
e_1)]$. Then by evaluating the lattice Green's function for an
infinite square \cite{SPITZER}, one finds $\phi(\vec
0)=\rho/[1-\epsilon/4]$.

To determine the large distance behavior of the solution, one can use the continuum approximation under which
the Green's function behaves as $G(\vec r, {\vec r_0})\approx
-\frac{1}{2\pi} \ln|\vec r-{\vec r_0}|$, for large $|\vec r-{\vec
r_0}|$. Substituting in \eqref{dens1}, one finds that the density in \eqref{dens1} decays for
large $r$ algebraically as
\begin{equation}
\phi(\vec r) = \rho- \frac{\epsilon \phi(\vec 0)}{2\pi}
\frac{{\vec e_1}\cdot {\vec r}}{r^2} +
\mathcal{O}(\frac{1}{r^{2}}).
\label{dipole1}
\end{equation}
This density profile also leads to a nontrivial current profile. The
average particle current density, away from the drive, is ${\vec j}(\vec
r)=-\nabla
\phi(\vec r)$ and it decays for large $r$ as
\begin{equation}
{\vec j}(\vec r)= \frac{\epsilon \phi(\vec 0)}{2\pi}\, \frac{1}{r^2}\, \left[{\vec e_1}-\frac{2({\vec e_1}\cdot {\vec r}){\vec
r}}{r^2}\right] +
\mathcal{O}(\frac{1}{r^{3}}).
\label{curr1}
\end{equation}
In the electrostatic analogue,
${\vec j}(\vec r)$ is precisely the electric field generated by the
dipole.

Due to the superposition principle, the above analysis can be
readily generalized to the case of arbitrary localized configuration
of the driving bonds. For example, consider a case of two driving bonds
with rates $(1-\epsilon)$ each, one from $(0,0)$ to $(1,0)$ and the
other from $(0,0)$ to $(-1,0)$, while the rates across the rest of
the bonds in the lattice are fixed to be $1$ in both ways:
\begin{equation}
\cdots (-2,0)\, \overset{1}{\underset{1}{\leftrightarrows}}\,
(-1,0)\, \overset{1-\epsilon}{\underset{1}{\leftrightarrows}}\,(0,0)\,
\overset{1-\epsilon}{\underset{1}{\rightleftarrows}}\, (1,0)\, \overset{1}{\underset{1}{\rightleftarrows}}\,
(2,0)\cdots \nonumber
\end{equation}
It is again easy to see that the steady state density
$\phi(\vec r)$ now satisfies
\begin{equation}
\nabla^2 \phi(\vec r) =-\epsilon \phi(\vec 0)\left[2\delta_{\vec r,\vec 0}-\delta_{\vec r,\vec e_1}-\delta_{\vec r, -\vec e_1} \right].
\label{quad1}
\end{equation}
In the electrostatic analogue, the rhs of \eqref{quad1} corresponds
to two oppositely oriented adjacent dipoles on the $x$ axis,
constituting a quadrupole charge configuration $(-++-)$ in the
continuum limit. Using the two-dimensional Coulomb potential and the
superposition principle, it is easy to see that the density profile
at large distance $r$ now decays as
\begin{equation}
\phi(\vec r) = \rho-\frac{\epsilon\phi(\vec
0)}{2\pi}\left[\frac{1}{r^{2}}-2\left(\frac{{\vec e_1}\cdot {\vec
r}}{r^2}\right)^2\right]+ \mathcal{O}(\frac{1}{r^{4}}),
\label{quadrupole1}
\end{equation}
with $\phi(\vec{0})=\rho/(1-\epsilon/2)$.
Consequently, the particle current density (or equivalently the
electric field of the quadruple) ${\vec j}(\vec r)$ decays as
$r^{-3}$ for large $r$.

In the case of an arbitrary configuration of $n$ driving bonds, one
uses the superposition principle to express the steady state profile
in terms of the dipole strengths of the $n$ dipoles. Using the exact
expression for the Green's function on square lattice \cite{SPITZER}
a set of linear equations determining these strengths are obtained,
which can be readily solved.

The existence of biased bonds does not necessarily imply a breakdown
of detailed balance. Localized configurations of biased bonds may
preserve detailed balance with respect to a localized potential
$V(\vec r)$. For example, consider the case where all incoming links
to the site $(0,0)$ are with rates $(1-\epsilon)$ each and the
rates across the rest of the links in the lattice are fixed to be
$1$. It is easy to verify that the rates satisfy
detailed balance with respect to the localized potential $V(\vec r)=
-\ln(1-\epsilon)\, \delta_{\vec r, \vec 0}$. Consequently, the
steady state density has the Gibbs-Boltzmann form $\phi(\vec r)
\propto \exp[-V(\vec r)]$ leading to a flat density profile
everywhere except at the origin.

It is interesting to note that in $d=2$ dimension, the analogy to an
electrostatic problem can be extended to introduce a magnetic field
as well. In a general two-dimensional setting let $e_{ij}$ denote
the hopping rate from site $i$ to its nearest neighbor site $j$.
This link creates a pair of oppositely directed flux lines centered
on the two plaquettes that share this link. The flux is perpendicular
to the plaquettes,
with, say, the flux to the left of the link points up (see Fig.\ref{fig1}a). The magnitude of the field generated by the $(ij)$
bond is given by $H=\ln\left(\frac{e_{ij}}{e_{ji}}\right)$. The
total flux through each plaquette is given by the sum of the fluxes
generated by each of its links. A necessary and sufficient condition
for detailed balance to hold is the vanishing of the total magnetic
field, as defined above, on all plaquettes. This is a direct
consequence of the Kolmogorov criterion \cite{KLMGRV,DAVID}. In that
case, the density profile outside the driven region is flat resulting
in a vanishing electric field, as follows from the Boltzmann
measure. However when the magnetic field is non-zero, the steady
state is a nonequilibrium one, and the density profile created by the
driving bonds typically decays algebraically. On the other hand, bond configurations
which result in vanishing electric field and non-vanishing magnetic
field has a flat density profile. An example of such configuration
is given in Fig.\ref{fig1}b.

Another interesting density profile pattern emerges when one applies
a global bias, say in the $x$-direction. Consider again
noninteracting particles in $d=2$, where a particle from any site
$(m,n)$ hops to a neighboring site with rate $1$ in the north, west
and south directions, while with rate $(1+\mu)$ to the eastern
neighbor. Thus, $\mu\ge 0$ denotes the global bias. In addition,
there is the driving bond from $(0,0)\to (1,0)$ where the hopping rate
is set to be $(1+\mu-\epsilon)$ with $0\le \epsilon\le 1$.
Proceeding as before, in the steady state, $\phi(\vec r)$ is found to
satisfy
\begin{equation}
-\mu\,\left[\phi(\vec r)-\phi(\vec r-\vec e_1)\right]+ \nabla^2 \phi(\vec r) =-\epsilon \phi(\vec 0)\left[\delta_{\vec r,\vec 0}-\delta_{\vec
r,\vec e_1}\right]
\label{bias1}
\end{equation}
where $\vec r-\vec e_1 \equiv (m-1,n)$ and $\nabla^2$ is the discrete Laplacian as before. The solution can be expressed
as
\begin{equation}
\phi(\vec r)= \rho +\epsilon \phi(\vec 0) \left[g(\vec r, \vec 0)-g(\vec r, \vec e_1)\right]
\label{biassol1}
\end{equation}
where the Green's function $g(\vec r, \vec r_0)$ satisfies
\begin{equation}
-\mu\,\left[g(\vec r, \vec r_0)-g(\vec r-\vec e_1, \vec r_0)\right]+ \nabla^2 g(\vec r)=-\delta_{\vec r, \vec r_0}.
\label{biasgreen1}
\end{equation}
The solution can be obtained using Fourier transformation. Setting
$(X,Y)\equiv \vec r-\vec r_0$, one finds that
for large $X$, $Y$
\begin{eqnarray}
g(X,Y) &\simeq & \frac{1}{\sqrt{4\pi \mu\, X}}\, e^{-\mu Y^2/{4X}}, \,\quad X> 0 \label{posX}\\
& \simeq & \frac{e^{\mu X}}{\sqrt{4\pi \mu\, |X|}}\,   e^{-\mu Y^2/{4|X|}},\,\quad X< 0 \label{negX}.
\end{eqnarray}
These results have a nice interpretation as the solution of a
diffusion equation where $X$ plays the role of `time' and $Y$ the
distance traveled from the origin. It can be directly seen from
\eqref{biasgreen1} which, for large $(X,Y)$, can be approximated by
its continuum version: $-\mu \partial_X g + \partial_X^2 g+
\partial_Y^2g=0$. For large $X>0$, neglecting the term
$\partial_X^2$, one indeed obtains an analogue of diffusion
equation, $\mu \partial_X g= \partial_Y^2 g$ with $\mu$ playing the
role of friction coefficient and $X>0$ being the `time' variable and
hence one obtains the standard diffusive propagator in \eqref{posX}
for $X>0$. In contrast, for negative $X$, one can no longer
interpret directly in terms of the diffusion equation in which the
`time' is always a positive variable. However, upon making a change
of variable $X\to -X'$ and substituting $g(X,Y)= e^{-\mu X'} h(X',Y)$ one gets
$\mu \partial_{X'} h=\partial_{X'}^2 h+ \partial_Y^2 h$.
This leads to the result \eqref{negX} for $X<0$. Substituting this
result for the Green's function in \eqref{biassol1} one obtains a
density profile $\phi(x,y)$ that is highly anisotropic. For example,
for $y=0$ and as $x\to +\infty$, the density
decays algebraically to $\rho$ as $x^{-3/2}$ while in the
direction opposite to the bias $x\to -\infty$, the density decays
exponentially. In the $y$ direction, for fixed $x$, the density
falls off rapidly as $\sim \exp\left[-\mu y^2/{4|x|}\right]$.

The $2$-d results obtained above for noninteracting particles, with
or without global bias, can be
easily generalized to arbitrary dimensions. Indeed, the solution in
\eqref{dens1} holds for arbitrary dimensions $d$, except that the
Green's function $G(\vec r, \vec r_0)$ depends on $d$. In the
continuum limit, the Couloumb potential $G(\vec r, \vec r_0)$
behaves, for large $|\vec r-\vec r_0|$, as $|\vec r-\vec
r_0|^{-(d-2)}$ for $d>2$, as $ -\frac{1}{2\pi}\ln|\vec r-\vec r_0|$
for $d=2$ and as $ -\frac{1}{2}|\vec r-\vec r_0|$ for $d=1$. Hence,
the dipole potential and consequently the density profile $\phi(\vec
r)$ decays as $ r^{-(d-1)}$ for large $r$ in $d\ge 2$. In $d=1$,
the dipole potential has a discontinuity at $x=0$, thus giving rise
to a discontinuous density profile: $\phi(x)= \rho-(\epsilon/2)\,
\phi(0)\, {\rm sgn}(x)$, in full accordance with \cite{BDL}. The results for
quadrupoles and higher
multi-poles can similarly be generalized to arbitrary dimensions.
The analogy to a magnetic
field discussed earlier is, however, restricted only to two dimensions.
\begin{figure}
\includegraphics[width=1.0\hsize]{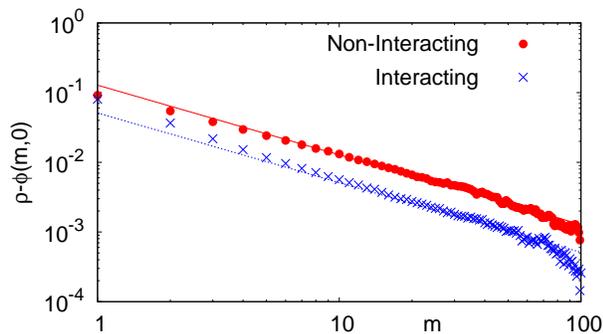}
\caption{The algebraic decay of the density profile away from the
driving bond in the positive $x$ direction.}
\label{fig3}
\end{figure}

We now show that most of the results derived above for
noninteracting particles in presence of a localized drive carry
through when the hard core interaction between the particles is
switched on. We consider a symmetric exclusion process on a $2$-d
square lattice where each site can hold at most one particle. Our results are easily generalizable to
arbitrary dimensions. From
any occupied site $(m,n)$ the particle attempts to hop to any of its
neighboring sites with rate $1$ and actually hops there provided
the target site is empty. As in the noninteracting case, we
introduce the localized drive across the bond $(0,0)\to (1,0)$ where
the attempted hopping rate is $(1-\epsilon)$ with $\epsilon \le
1$. It is useful to first associate an occupation variable
$\tau(\vec r,t)$ with every site $\vec r$: $\tau(\vec r,t)=1$ if the
site $\vec r$ is occupied at time $t$ and is zero if it is empty at
$t$. Clearly, the average density is $\phi(\vec r,t)= \langle
\tau(\vec r,t)\rangle$. It is again easy to write down the time
evolution equation of $\phi(\vec r,t)$ by counting the incoming and
outgoing rates from each site.
In this case the equation analogous to \eqref{diff1} is
\begin{equation}
\nabla^2 \phi(\vec r)=-\epsilon \langle \tau(\vec 0)(1-\tau(\vec e_1)\rangle\, \left[\delta_{\vec r,\vec 0}-\delta_{\vec r,\vec e_1}\right],
\label{sep1}
\end{equation}
where $\phi(\vec 0)$ in \eqref{diff1} is replaced by
$C=\langle \tau(\vec 0)(1-\tau(\vec e_1)\rangle$.
While, unlike in the noninteracting case, we can not determine this
prefactor self-consistently, the electrostatic analogy to a dipole
(albeit with an unknown strength $\epsilon C$) still holds. Thus, at
long distances, one still obtains a long-ranged algebraic decay of
the density profile
\begin{equation}
\phi(\vec r)= \rho -  \frac{\epsilon C}{2\pi} \frac{{\vec
e_1}\cdot {\vec r}}{r^2}+\mathcal{O}(\frac{1}{r^{2}}).
\label{sepdens1}
\end{equation}
Consequently, the particle current density (equivalently the
electric field due to the dipole) ${\vec j}(\vec r)$, decays at
large distances as in the noninteracting case \eqref{curr1}, up to
an overall multiplicative constant. In a similar way, one can also
arrange the driving bonds so as to generate a quadrupole or
multi-pole configurations of charges giving rise to an algebraic
decay of the density profile with varying exponents depending on the
charge configurations.

A numerical evidence of the profiles in (\ref{dipole1}) and (\ref{sepdens1}) is
shown in Fig.2 where the difference of density from
$\rho$ is plotted against the distance from the driving bond
in the positive $x$ direction. The simulation is
performed on a $200\times200$ lattice with $\epsilon=1$ and initial uniform
density $\rho=0.6$. The straight lines denote the
theoretical results in (\ref{dipole1}) and (\ref{sepdens1}) with
the value of $\phi(0)$ calculated using $\rho$ and $\epsilon$, and
$C=\langle\tau(0)(1-\tau(\vec{e}_{1}))\rangle=0.3209$
determined independently from the Monte Carlo simulation.

In summary, we have demonstrated that in diffusive systems, both with and without
inter particle exclusion interaction, localized drive can give rise to
algebraically decaying density profiles at large distances.
The problem of determining the density profile is mapped onto an
electrostatic problem where each driving bond is represented by an
electric dipole whose strength is determined self-consistently by the
electric potential generated on the driving bond. The density profile
of the driven system is then given by the electrostatic potential created by
the charge distribution. An
analogous quantity to the magnetic field is also identified in two
dimensions.

We thank M. R. Evans and O. Hirschberg for their constructive comments on an earlier
draft of this paper. The support of the Israel Science Foundation
(ISF) is gratefully acknowledged. This work was carried out while
S.N.M. was a Weston Visiting Professor at the Weizmann Institute of
Science.

\end{document}